\documentclass{article}
\usepackage{icmc2026_paper_template}
\usepackage{times}
\usepackage{ifpdf}
\usepackage{soul}
\usepackage[english]{babel}

\usepackage{amsmath} % popular packages from Am. Math. Soc. Please use the 
\usepackage{amssymb} % related math environments (split, subequation, cases,
\usepackage{cite} % Allows for more control over ref formatting 

\def\papertitle{Extending Xenakis: From Architectural Geometry to Sonification of the Philips Pavilion}
\def\firstauthor{First Author}
\def\secondauthor{Second Author}
\def\thirdauthor{Third Author}

\newif\ifpdf
\ifx\pdfoutput\relax
\else
   \ifcase\pdfoutput
      \pdffalse
   \else
      \pdftrue
  \fi
\fi

\ifpdf % compiling with pdflatex
  \usepackage[pdftex,
    pdftitle={\papertitle},
    pdfauthor={\firstauthor, \secondauthor, \thirdauthor},
    bookmarksnumbered, % use section numbers with bookmarks
    pdfstartview=XYZ % start with zoom=100% instead of full screen; 
   ]{hyperref}
  \usepackage[pdftex]{graphicx}
  \graphicspath{{./figures/}}
  \DeclareGraphicsExtensions{.pdf,.jpeg,.png}

  \usepackage[figure,table]{hypcap}

\else % compiling with latex
  \usepackage[dvips,
    bookmarksnumbered, % use section numbers with bookmarks
    pdfstartview=XYZ % start with zoom=100% instead of full screen
  ]{hyperref}  % hyperrefs are active in the pdf file after conversion

  \usepackage[dvips]{epsfig,graphicx}
  \graphicspath{{./figures/}}
  \DeclareGraphicsExtensions{.eps}

  \usepackage[figure,table]{hypcap}
\fi

\hypersetup{
    colorlinks,%
    citecolor=black,%
    filecolor=black,%
    linkcolor=black,%
    urlcolor=black
}

\title{\papertitle}

\fourauthors
{1.5in}
{Changda Ma} {Georgia Institute of Technology \\ %
{\tt \href{cma326@gatech.edu}{cma326@gatech.edu}}}
{Sunshiyu Wang} {Georgia Institute of Technology \\ %
{\tt \href{swang3214@gatech.edu}{swang3214@gatech.edu}}}
{Canting Zhu} {Georgia Institute of Technology \\ %
{\tt \href{czhu348@gatech.edu}{czhu348@gatech.edu}}}
{Alexandria Smith} {Georgia Institute of Technology\\ %
{\tt \href{alexandria.smith@gatech.edu}{alexandria.smith@gatech.edu}}}

\begin{document}
\capstartfalse
\maketitle
\capstarttrue
\begin{abstract}
Architecture and music have been linked through proportion and temporal structure, yet architectural geometry is rarely viewed as a source of generative music. Revisiting Xenakis’ one-directional transformation from string glissandi in \textit{Metastaseis} to the ruled surfaces of the \textit{Philips Pavilion}, we invert this workflow and sonify the completed Pavilion as a temporal composition. We reconstruct the Pavilion as nine ruled surfaces, extract their governing ruling lines, and subdivide each surface into structural lines and spatial sampling points. Four evenly spaced ruling lines per surface generate continuous string glissandi, while 3{,}357 sampled points develop five density-based energy blocks and a sparse brass and woodwind subsequence. Implemented in Python, the system produces MIDI rendered in Ableton Live, accompanied by a real-time 3D visualization that reveals architectural motion, stasis, and structural contrast through sound and image. In general, this work paves the way for the transfer of architectural geometry as a performable musical structure, extending Xenakis’s architectural and musical thinking to sonification and interactive music practice.
\end{abstract}

\section{Introduction}\label{sec:introduction}
``Music is liquid architecture; architecture is frozen music.''
This famous phrase, attributed to Johann Wolfgang von Goethe, describes a structural analogy between architecture and music \cite{howat1977debussy}.
However, contemporary architectural practice reflects architecture as a medium for transmitting music, rather than as a generator of musical structure. 
Few studies investigate how architectural geometry itself can be translated into musical logic.

A foundational exploration linking architecture and music at the level of form generation traces back to the work of Iannis Xenakis.
In \textit{Metastaseis}, Xenakis projected the mathematical hyperbolic paraboloid onto the string glissandi using a graphic notation of time and pitch \cite{xenakis1967metastaseis}.
Therefore, the slope, continuity, and density of lines in the hyperbolic paraboloid projection determine the pitch change, continuity, and ensemble density \cite{wannamaker2012mathematics}.
Together with the modernist architect Le Corbusier, Xenakis translated music composed of glissando patterns into architectural form in the \textit{Philips Pavilion} \cite{sikiaridi2003architectures, xenakis1958philips}, where the musical structure informed the architectural form. 
However, the translation remained unidirectional: the architecture appeared as a static outcome rather than an active musical agent.

Recent advances in architectural parametric design \cite{hudson2010strategies} allow architectural elements such as surfaces, vertices, and ruled geometries to be accessed as structured and data-driven representations rather than fixed objects \cite{jabi2013parametric,tessmann2019geometry}.
This transformation makes architectural geometry suitable for sonification.
Sonification, as a method for translating data into non--speech sound \cite{walker2011theory}, provides a new acoustic dimension for architectural parametric design.

This study explores the sonification of architecture by reversing the historical process established by Xenakis.
Rather than reproducing Xenakis’s instrumentation, our approach takes the completed \textit{Philips Pavilion} and maps its geometry back into sound.
Using parametric architectural modeling, we extract the ruled-surface geometry of the \textit{Philips Pavilion} as spatial data for sonification. We map ruled lines to continuous string glissandi, use variations in point density to define energy-based blocks of stasis, and employ selected sampling points to generate the structure through sparse brass counterpoints.
By mapping architectural geometry into sound, this work reframes architecture as an active generative instrument and extends Xenakis’s architectural–musical thinking through contemporary computational methods.

\section{Background and Literature Review}\label{sec:Background and Literature Review}
In the eighteenth century, Gotthold Ephraim Lessing distinguished architecture as a spatial art and music as a temporal art \cite{speidel2020telling}.
While early twentieth-century movements and Le Corbusier’s concepts challenged this separation by introducing narrative or perceptual time \cite{brown2005physical, cutting2002representing, giedion2009space}, architectural space remained fundamentally static.
Even as composers such as Stockhausen explored the spatial dynamics of sound \cite{heathcote2003liberating}, architecture continued to serve primarily as a passive container, lacking generative feedback from musical organization.
   \begin{figure*}[ht!]
        \centering
        \includegraphics[width=2.0\columnwidth]{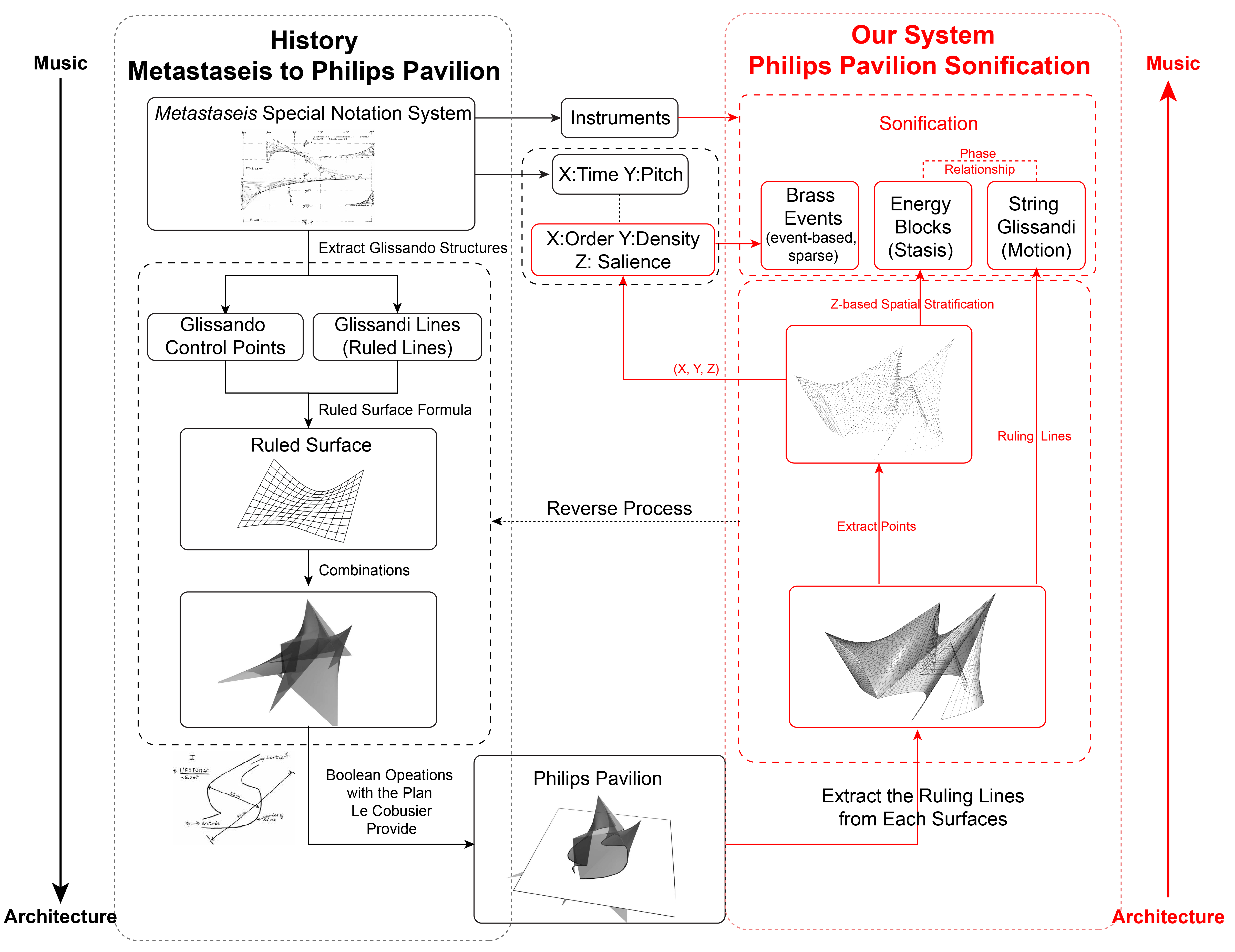}
        \caption{System workflow illustrating the inversion of Xenakis’ historical music to architecture process.}
        \label{fig:overflow}
    \end{figure*}
Xenakis’s practice achieved a breakthrough in the relationship between music and architecture.
By transferring the logic of musical structure to architectural form, he used mathematics to generate the geometric form of the \textit{Philips Pavilion} \cite{sterken2007music}. Rigorous mathematical logic was first used to establish a bridge between musical structure and architectural design principles \cite{clarke2012iannis}.
However, due to the absence of computational tools for extracting architectural geometry as quantifiable data, this relationship could not be operated in reverse.
Architectural form remained an endpoint rather than a generative source of music.

Despite this methodological limitation, the theoretical discourse has recognized the bi-direction of Xenakis’s vision. Beyond Anglophone literature, French scholarship emphasizes the reciprocal dialog inherent in Xenakis's interdisciplinary practice. Studies by Nathalie Boulouch \cite{boulouch2020severine} and Makis Solomos \cite{solomos2006diatope} frame the \textit{Philips Pavilion} not only as a static artifact derived from music, but as a disciplinary crossing where sonic and architectural thought remain entangled. 

Although advances in computational design have made architectural geometry available as data, few studies have explored architectural geometry itself as a generative source for sonification. Several studies have drawn inspiration from Xenakis to sonify architectural or urban forms.

Parthenios \cite{PanagiotisParthenios} adopted Xenakis’s \textit{Metastaseis}–to–\textit{Philips Pavilion} music-architecture mapping strategy to sonify architectural facades in urban contexts.
Although spatial dimensions and surface density were assigned to musical parameters, this research used correspondences without preserving the historical and structural logic from the architectural–musical synthesis of Xenakis.

Torresan et al. \cite{torresan2022singing} demonstrated how structural vibration data can be sonified by translating measured dynamic responses, such as acceleration and modal behavior, into sound.
However, this approach represents a system-response sonification based on engineering data rather than a sonification of the architectural geometric structure itself.

In contrast, research on the sonification of architectural geometric structures remains relatively scarce.
This gap motivates our system, which implements ruled-surface lines and point distributions as musical control structures.
\section{System Design}\label{sec:system design}
\subsection{History-based System Overview}\label{sec:ove}
Drawing inspiration from Xenakis’s compositional approach in \textit{Metastaseis} and its architectural realization in the \textit{Philips Pavilion}, our system revisits this historical transformation between sound and space.
In \textit{Metastaseis}, Xenakis used the distinctive glissando notation system to extract the control points and projection lines of the time-pitch traces for each glissando.
To transform these linear sounds into the design logic of a spatial surface, he applied the mathematical principles of the ruled surface, which was shown in Formula~\ref{eq:ruled_surface}. A ruled surface is formed by a family of geriatrics sweeping along a directrix \cite{krivoshapko2023kinematic}. Thus, surfaces can be generated entirely by lines.
\begin{equation}
S(u,v) = (1 - v)\,\boldsymbol{r}_1(u) + v\,\boldsymbol{r}_2(u), 
\qquad u,v \in [0,1]
\label{eq:ruled_surface}
\end{equation}
where $\boldsymbol{r}_1(u)$ and $\boldsymbol{r}_2(u)$ are the two ruling curves, and linear interpolation between them defines the surface geometry.

Through the selection and organization of glissandi with varying directional tendencies, Xenakis constructed multiple ruled surfaces that together defined the spatial and structural logic of the \textit{Philips Pavilion} \cite{clarke2012iannis}.
This process establishes linear trajectories as a shared structural unit between music and architecture.
Building on this principle, our system reverses this historical workflow by treating the completed architectural geometry of the \textit{Philips Pavilion} as a generative source for musical structure, using extracted lines and point distributions to drive motion, stasis, and event-based behaviors in sound.
\subsection{System Implementation: Extending Xenakis--Inverting the History Process}\label{sec:ove}
We adopt the principles through which \textit{Metastaseis} was transformed into the \textit{Philips Pavilion} and invert this workflow to design our ``Extending Xenakis'' system. 
From the history to our system, the full workflow of our system is illustrated in Figure~\ref{fig:overflow}.

Based on the ruled surface formula, we deconstruct the nine ruled surfaces of the Pavilion by extracting their ruling lines and applying parametric sampling points along these lines.
The spatial points \((x, y, z)\) generated by this procedure were used as data sources for our sonification process.
Although \textit{Metastaseis} was processed within a two-dimensional time--pitch framework, its logic lies not in a fixed coordinate system but in the continuous transformation of geometric form into sound.
Following this principle, we reinterpret the sampled spatial coordinates as a multidimensional control space for the musical structure, rather than as a direct parameter mapping. 
Within this framework, we extract ruling lines to generate continuous string glissandi that establish large-scale motion and convergence in pitch.
We organized the distributions of points, stratified along the vertical dimension, into energy-based blocks that represent moments of relative stasis following the completion of the glissandi.
In parallel, we select a reduced subset of points to generate sparse brass and woodwind events, forming a secondary, event-based layer that strengthens the structural endpoints of the string process.
By applying the same instruments used in \textit{Metastaseis}, the geometric form of the \textit{Philips Pavilion} was sonified as a temporal, audible inversion of the historical process through which Xenakis transformed music into architecture.\footnote{Code is available at: \url{https://github.com/WangSun725/Extending-Xenakis-Architecture-to-Midi-.git}}. 
\subsection{Architecture Geometry Processing}\label{sec:Architecture Geometry Processing}
\begin{figure}[h!]
    \centering
    \includegraphics[width=1\columnwidth]{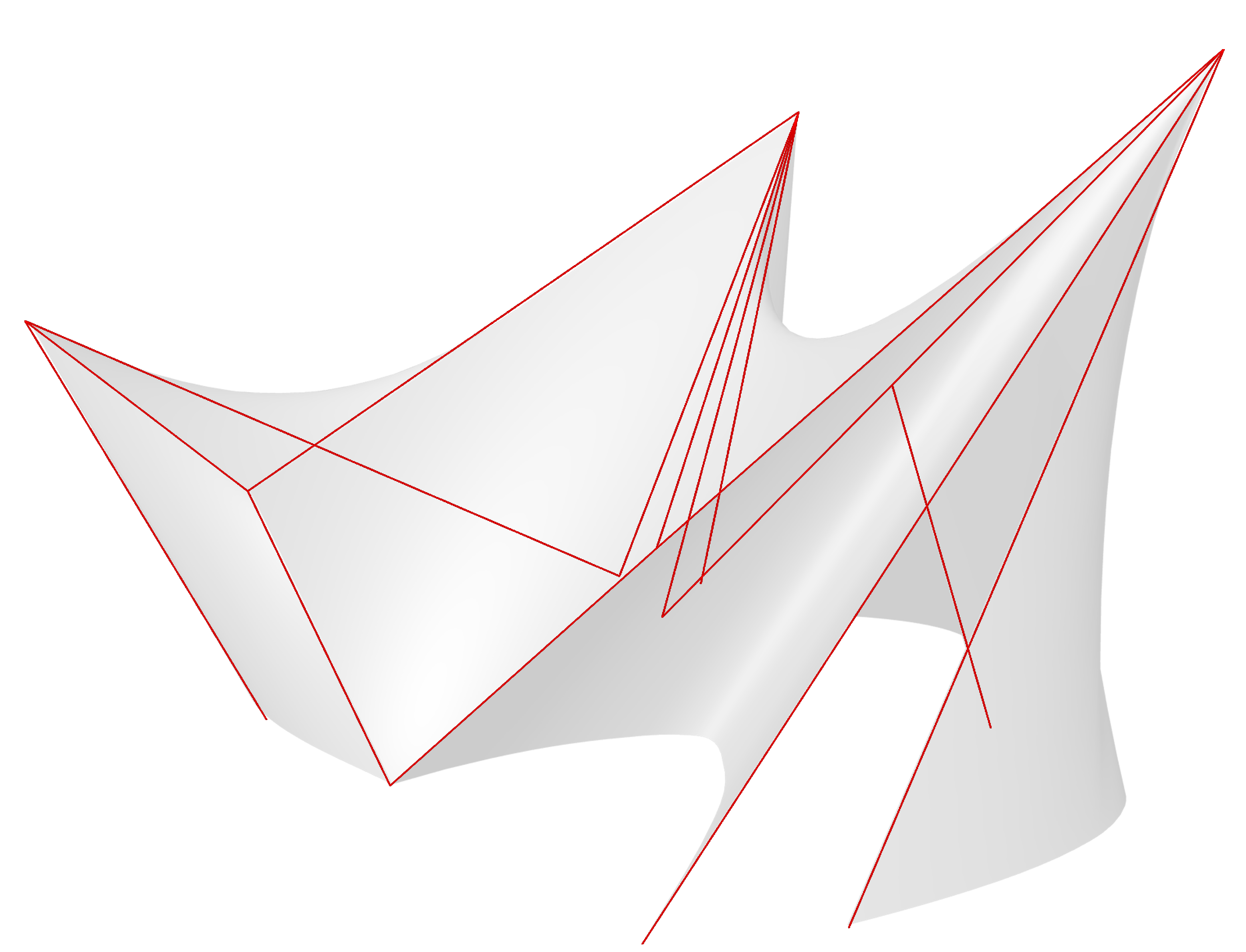}
    \caption{Ruling lines extracted from the Pavilion surface via the inverse ruled-surface process.}
    \label{fig:ruling lines}
\end{figure}

We reconstructed the geometry of the Philips Pavilion in Rhino based on archival drawings and historical documentation.
Using parametric modeling in Grasshopper, we rebuilt the \textit{Philips Pavilion}'s ruled surface system and extracted the nine ruled surfaces from the envelope.
For each given ruled surface $S(u,v)$, we recovered two governing ruling lines by the inversion process of Eq.~\ref{eq:inverse_ruled}, and Figure~\ref{fig:ruling lines} was shown as the result.
These ruling lines correspond to the generative trajectories that Xenakis associated with musical glissandi.
\begin{equation}
\left\{
\begin{aligned}
\mathbf{r}_1(u) &= S(u,0), \\
\mathbf{r}_2(u) &= S(u,1),
\end{aligned}
\right.
\qquad u \in [0,1].
\label{eq:inverse_ruled}
\end{equation}

\begin{figure}[h!]
    \centering
    \includegraphics[width=1\columnwidth]{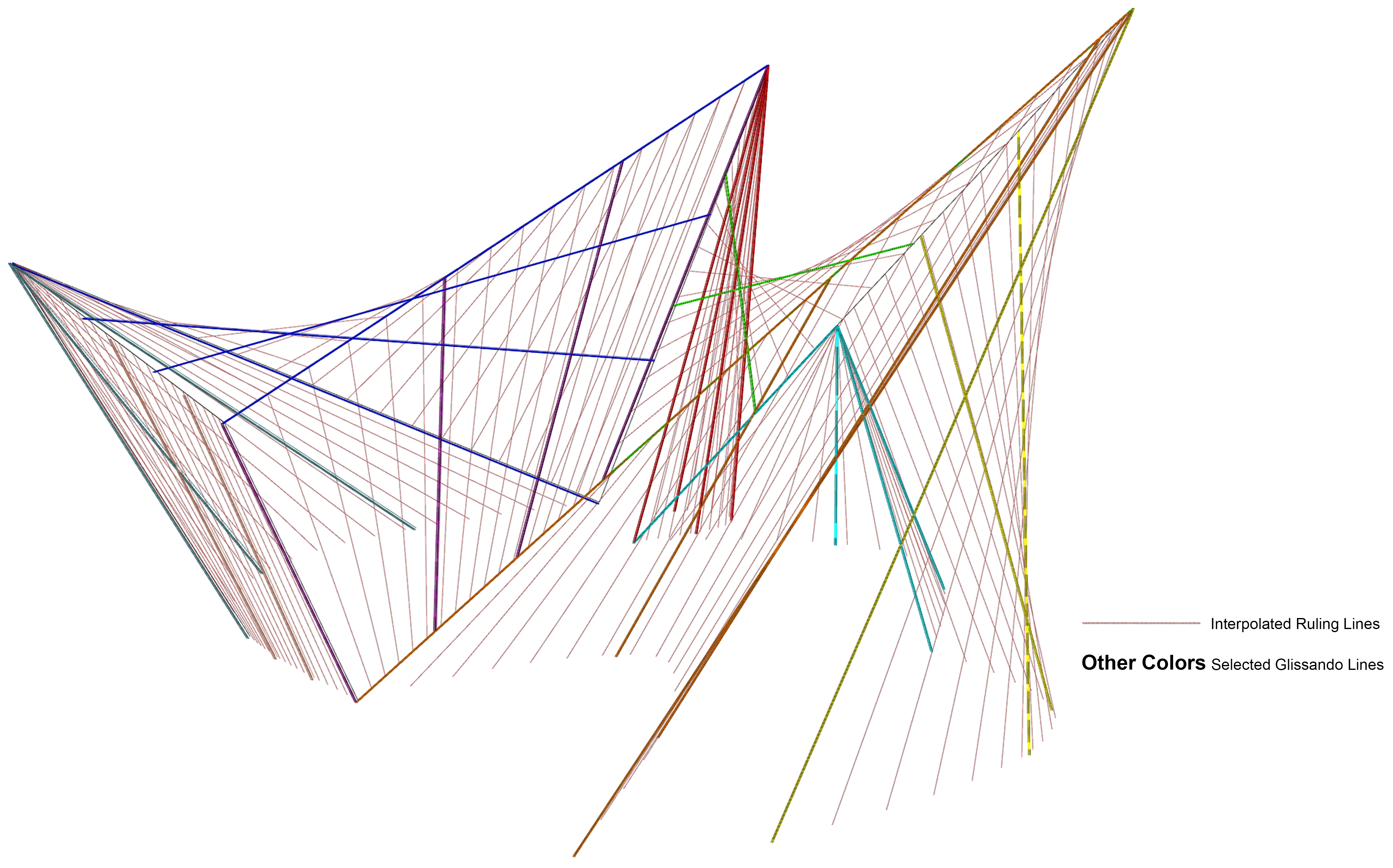}
    \caption{Interpolated structural lines generated between the ruling lines of each surface.}
    \label{fig:structure lines}
\end{figure}

To capture continuous variations in curvature, orientation and spatial gradient, we first subdivided each ruled surface into 20 interpolated structural lines between its two governing ruling lines (Figure~\ref{fig:structure lines}).
This dense line sampling provides a high-resolution geometric representation of each surface and paves the way for subsequent sonification processes.

From these 20 structural lines, we selected four evenly spaced lines per surface as the geometric backbone of the string section, yielding a total of 36 structural lines across the Pavilion.
For each selected line, we extracted its start point, end point, and total length, which together define the path and duration of each string glissando, as shown in Figure~\ref{fig:structure lines}.
\begin{figure}[h!]
    \centering
    \includegraphics[width=1\columnwidth]{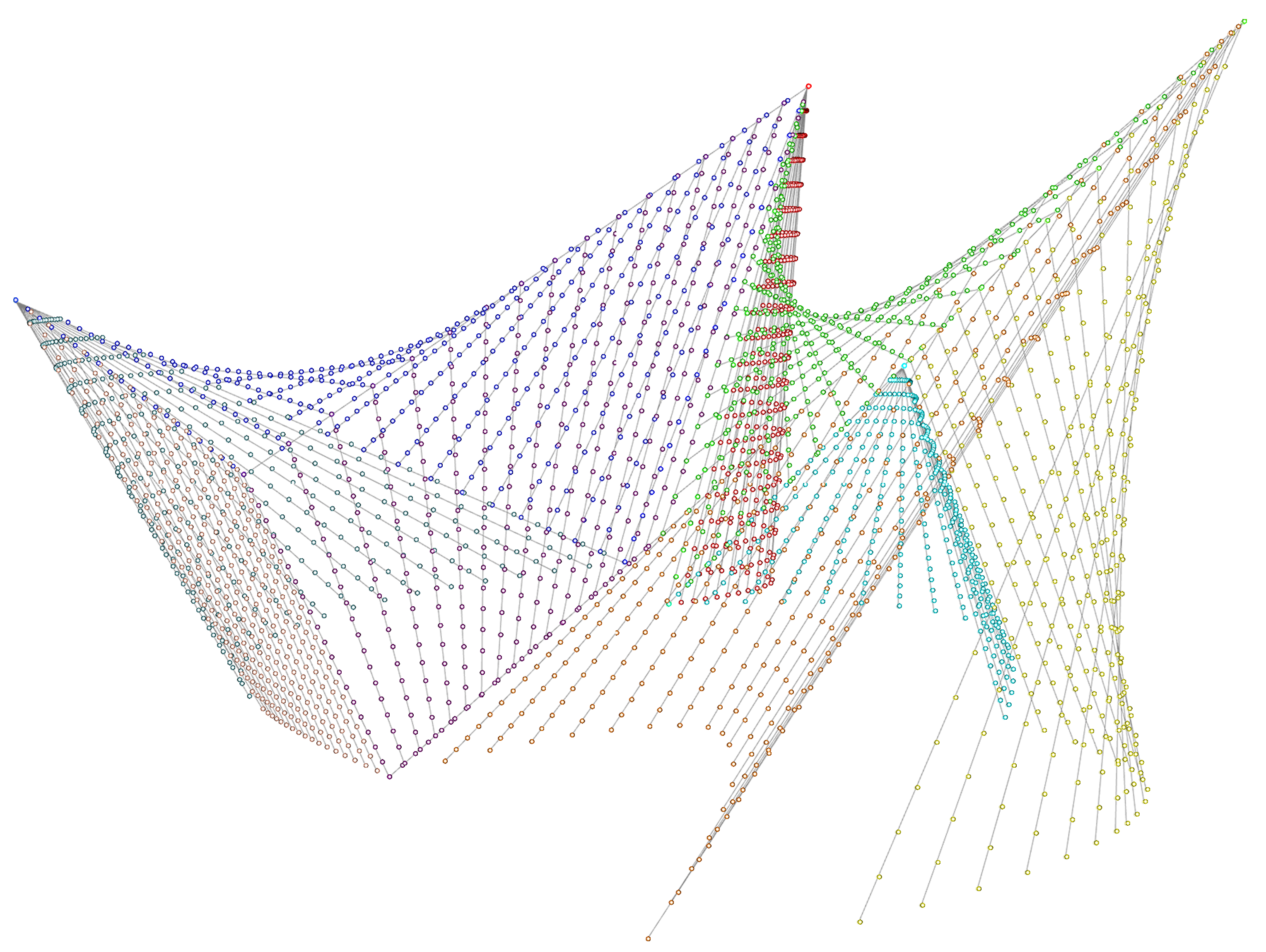}
    \caption{Spatial sampling points extracted from the Pavilion’s nine ruled surfaces, colored by surface groups.}
    \label{fig:sampling points}
\end{figure}

In addition, we sampled 20 evenly distributed spatial points along each structural line, yielding a total of 3,357 spatial points (Figure~\ref{fig:sampling points}) that form the input dataset for energy-based blocks and event-based brass sonification. 

Each sampled point was labeled according to its ruled surface, resulting in nine surface-based point groups.
Our system preserves geometric coherence across the sonification process and applies a consistent temporal organization derived from architectural form.

\subsection{Sonification Mapping}\label{sec:Sonification Mapping}

Our sonification system is designed both to generate a musical piece that  represents architectural characteristics and to mirror Xenakis’s architectural to musical logic. We create a set of algorithms to transform data extracted from architectural structures into a multi-layer MIDI composition organized in two musical phases. 

We organize the composition in a manner analogous to \textit{Metastaseis}, beginning with a continuous string glissando phase derived from ruled-surface line segments, followed by discrete energy blocks driven by spatial point distributions. We then select a reduced subset of sampling points to generate a brass and woodwind subsequence, mapping spatial positions to MIDI note events to enrich the musical texture.

The entire implementation is fully deterministic and is executed in Python, producing MIDI files that can be imported into Ableton Live for orchestration, positioning, mixing, and final rendering. 

\subsubsection{String Sonification} 
\label{sec:sonification-gliss} 

Our sonification output primarily focuses on string instruments. We divided the string materials into two sections: a ``Glissando'' section, followed by a ``Block'' section. In the ``Glissando'' section, we extract four line segments from each architectural surface, yielding a total of 36 lines. Each line segment is mapped to an independent string voice, resulting in an ensemble consisting of 12 violins, 8 violas, 8 cellos, and 8 double basses. We initialize all string voices in a common pitch class (G), adjusting the octave placement according to the pitch range of each instrument.

Each string instrument performs a sustained glissando, with MIDI pitch information derived directly from the geometric properties of its corresponding line segment. 
For each line segment, we compute its total geometric length from its start and end points in three-dimensional space.
We normalize line lengths and apply a nonlinear shaping strategy to preserve perceptual contrast and avoid extreme temporal values, as shown in Eq.~\ref{eq:duration_shaping}.
\begin{equation}
\hat{L}_i = \tilde{L}_i^{\gamma},
\label{eq:duration_shaping}
\end{equation}
where $\gamma > 1$ biases the distribution towards shorter durations. The final glissando duration $T_i$ is obtained by mapping $\hat{L}_i$ to a bounded temporal range, as shown in Eq.~\ref{eq:bound}.
\begin{equation}
T_i = T_{\min} + \hat{L}_i (T_{\max} - T_{\min}),
\label{eq:bound}
\end{equation}
where $T_{\min}$ and $T_{\max}$ define the minimum and maximum allowed glissando durations.

We render pitch evolution as a continuous glissando rather than as a sequence of discrete note steps.
For each line segment, we interpret the vertical displacement between its start and end points as a directional pitch tendency, scaled to the vertical extent of the Pavilion geometry.
This vertical variation determines the range and direction of pitch movement, allowing each string voice to trace paths through the pitch space.
We realize this pitch motion using continuous pitch bending, so that changes in height are perceived as uninterrupted glissandi rather than stepped intervals.
Throughout the duration of each glissando, the pitch is updated gradually and evenly over time, maintaining perceptual smoothness and preserving the continuous character of the string texture.

At the end of the ``Glissando'' section, each string voice maintains the final pitch reached at the end of its path. After a manually defined one second rest, the subsequent string ``Block'' section begins, and all string instruments remain at this final pitch and organize synchronized tremolo gestures, forming a sequence of energy blocks with coordinated onsets and offsets.

Block durations are derived from the architectural spatial point data. 
We divide the Pavilion geometry into five equal-width sections along the global vertical axis and assign each sampled point to a block index based on its $z$-coordinate. 
For each block, we compute the number of points as a proxy for the local structural density. These values are linearly mapped to block durations within a predefined temporal range. When all blocks contain equal numbers of points, a uniform midpoint duration is assigned.
The five blocks are arranged in order, separated by fixed rests, and all string voices end simultaneously at the end of the section.
\begin{figure*}[ht!]
\centering
\includegraphics[width=1.0\linewidth]{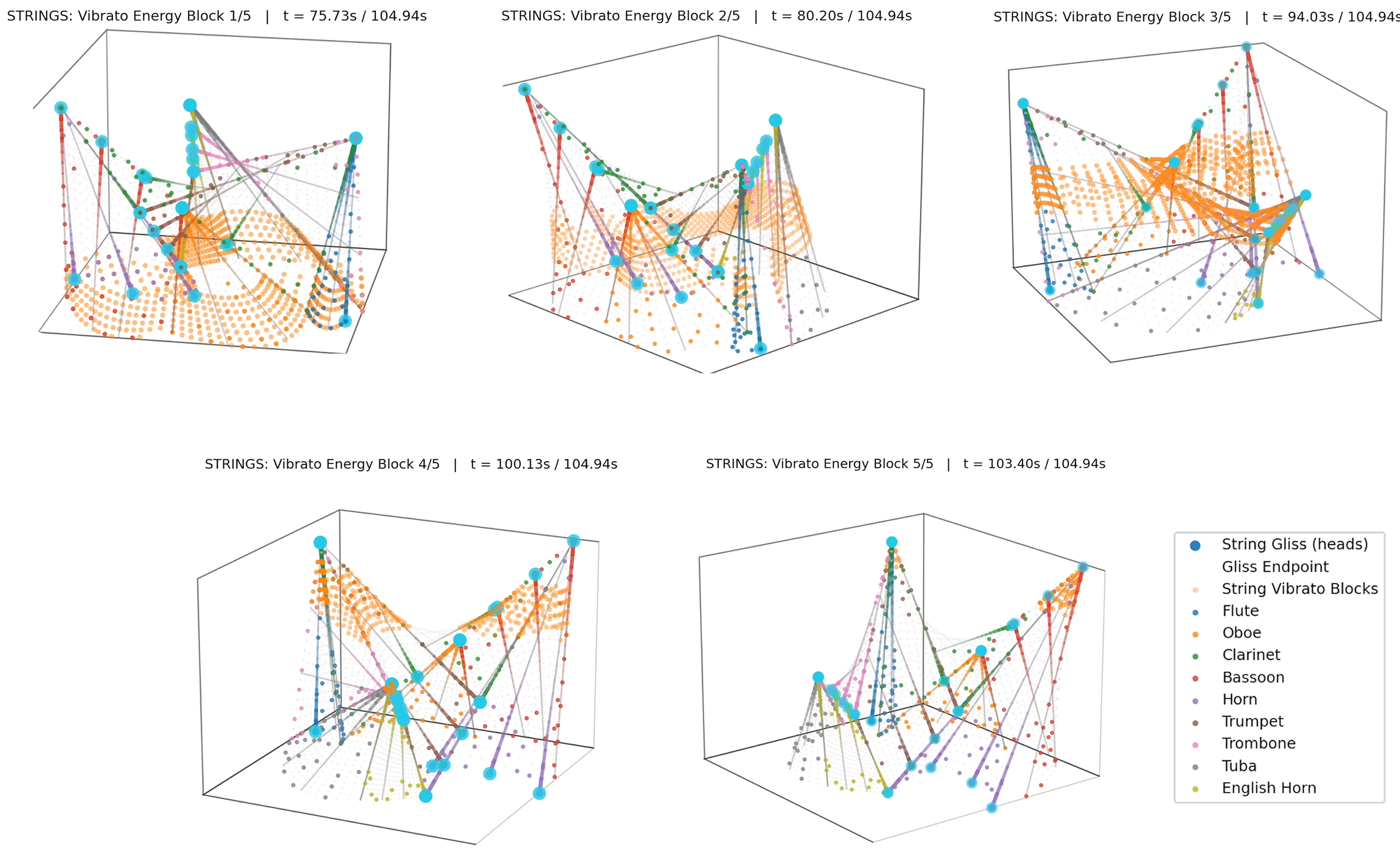}
\caption{Real-time visualization during the energy block phase.}
\label{fig:energy}
\end{figure*}
\subsubsection{Brass and Woodwind Subsequence}\label{sec:sonification-winds}

To enrich the musical texture and fully represent the architectural to musical transformation, we convert a subset of spatial points into nine MIDI tracks assigned to brass and woodwind instruments, forming a secondary structural layer synchronized with the string texture. Unlike the string section, which is derived from ruled lines and represents continuous geometric trajectories, the brass and woodwind layer is derived from discrete point samples and is designed to articulate localized structural moments as punctuated events rather than sustained motion. In this way, the two layers have a distinction presented in the architectural data: lines are rendered as continuity, whereas points are rendered as discreteness.

The brass and woodwind section consists of flute, oboe, clarinet, bassoon, horn, trumpet, trombone, tuba, and English horn. 
A reduced set of architectural points is sampled from the full point dataset and distributed across the nine instruments. This reduction is a structural control strategy that also remains consistent with the instrumental logic inherited from \textit{Metastaseis} \cite{read1976extending}. If the entire point dataset were converted into note events, the density of the piece would saturate the musical texture and obscure the distinction between the continuous, surface-based motion of the strings and the event-based articulation of the point layer. By limiting the number of active points, the brass and woodwind subsequences remain subordinate to the string section, while preserving the presence of discrete geometric features within the whole sonification. 

Each point is represented by a three-dimensional coordinate $(x_i, y_i, z_i)$ and mapped to a MIDI note event through a structured spatial to musical transformation. We use the vertical coordinate $z_i$ to encode pitch information by normalizing it relative to the global vertical extent of the dataset and mapping it to a bounded pitch range.
For brass and woodwind instruments, the resulting pitch values are discretized and quantized to a diatonic pitch set in G major, yielding discrete MIDI note numbers. 
This discrete pitch mapping distinguishes the brass and woodwind layers from the continuous pitch trajectories used in the string section.

We determine the temporal placement of each event from the horizontal coordinate $x_i$.
Each $x_i$ is linearly mapped to an onset time $t_i$ within the global timeline defined by the string section and quantized to a fixed rhythmic grid. 
The remaining coordinate $y_i$ modulates the event density and instrumental distribution rather than directly encoding the pitch or time. 
To maintain perceptual clarity, we encode a minimum inter-onset interval within each individual instrument track.

During the rest portions of the string section , we further thin brass and woodwind activity through probabilistic downsampling, reducing event density while preserving continuity across the full timeline.
Through these combined mappings and constraints, the brass and woodwind layer functions as a supporting and complementary component that punctuates and reinforces the structural endpoints of the overall sonification process.

\subsection{Real-Time Visualization}\label{sec:Real-Time Visualization}

We implemented a real-time sonification visualization developed in 
Python using matplotlib's 3D toolkit and FuncAnimation for real-time animation, with video export performed via FFMpegWriter \cite{hunter2007matplotlib}.

Rather than introducing an independent visual timeline, we drive all visual events from the same global time reference used in the sonification system.
Therefore, spatial activation, musical structure, and animation playback remain synchronized throughout the composition.

The visualization operates on two architectural datasets from the Pavilion geometry: ruled-surface line segments and spatial sampling points.
These datasets correspond to the string glissandi, energy blocks, and brass events introduced in our sonification system.

\begin{figure*}[ht!]
\centering
\includegraphics[width=1.0\linewidth]{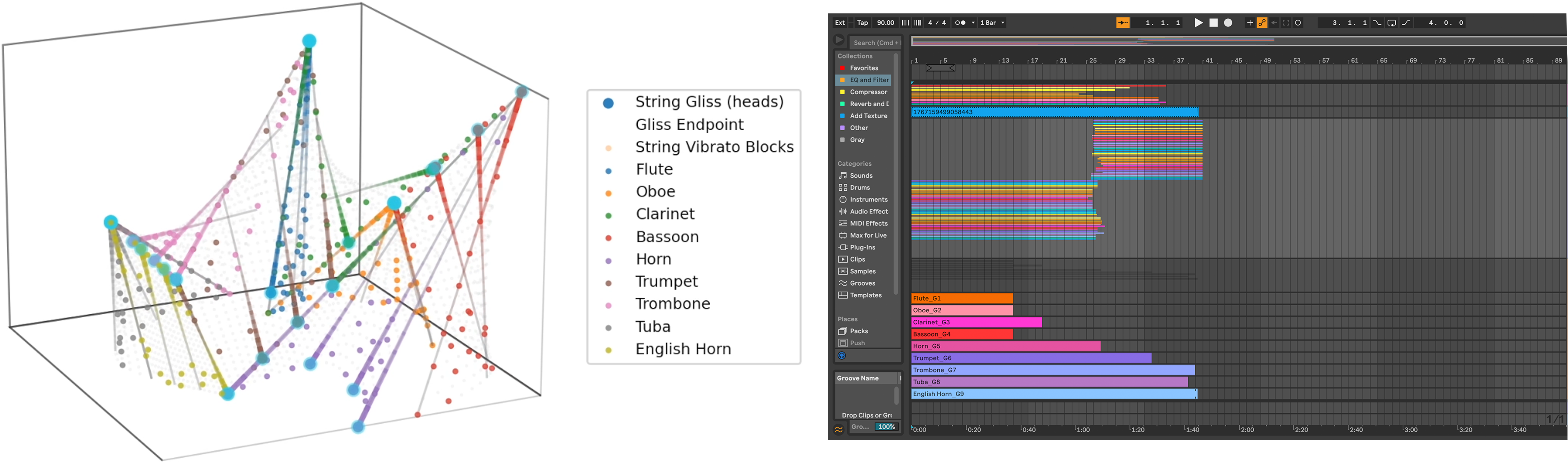}
\caption{Real-Time Sonification Visualization System}
\label{fig:geometry_animation}
\end{figure*}
Before the animation, the architectural geometry of the \textit{Philips Pavilion} is presented as a low-opacity, light-gray point cloud, together with 36 glissandi ruling lines shown in a muted, inactive state.
These background elements remain visible throughout the visualization, providing a spatial context and the form of the Pavilion.
Each ruled-surface group maintains a consistent color identity across phases, so the same architectural groups remain legible whether they are shown as lines, blocks, or event points.

\begin{figure}[ht!]
\centering
\includegraphics[width=1.0\linewidth]{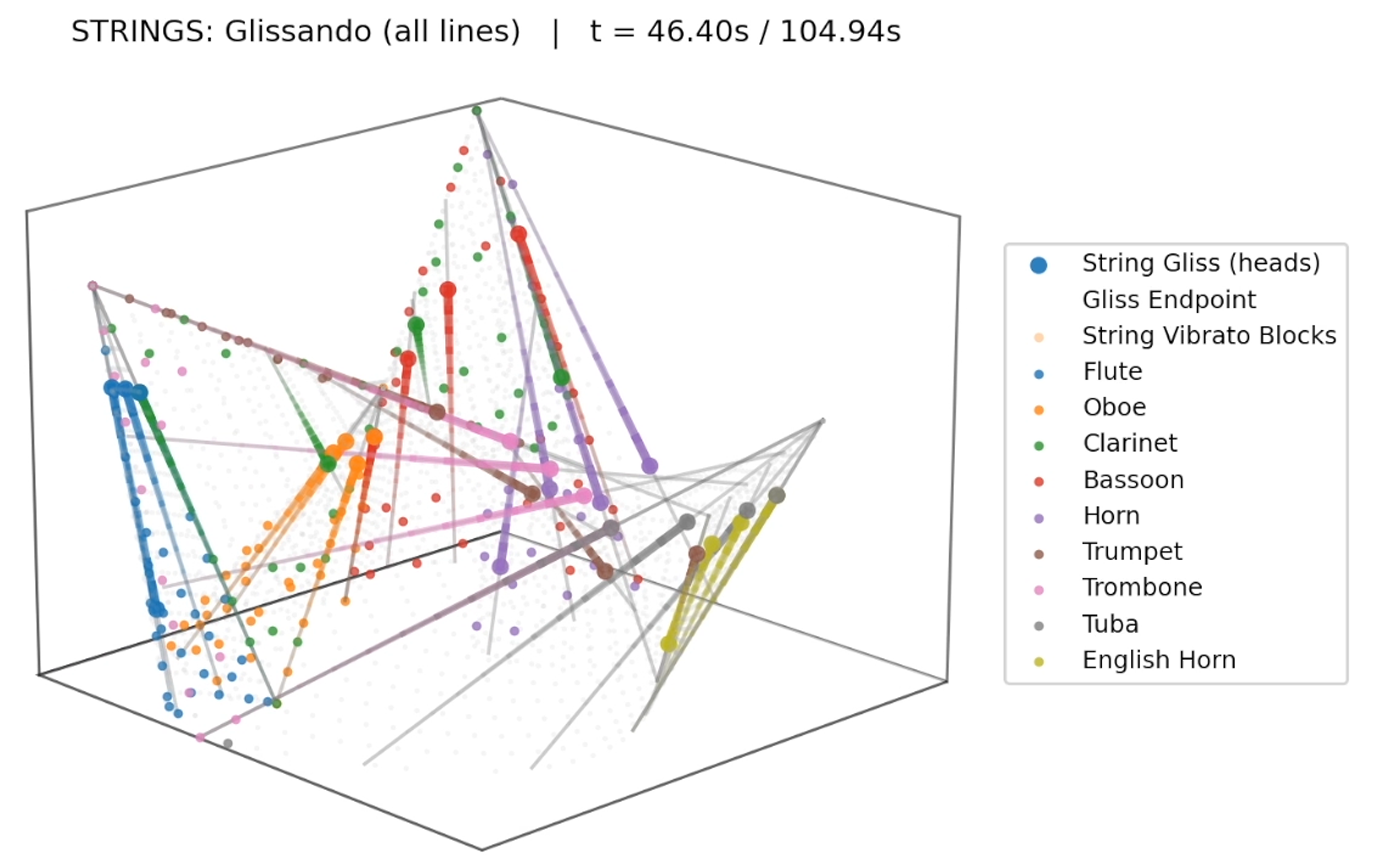}
\caption{Real-time visualization during the string glissando phase.}
\label{fig:glissando}
\end{figure}

During the opening animation, each string glissando is visualized as a colored traversal from the start point to the end point of its corresponding line segment.
A bright moving head and a short trailing trace make the direction and continuity of motion perceptible, while the gray scaffold remains a stable reference of the geometry.
In this phase, the endpoints are kept visually suppressed, so the attention is focused on the collective motion rather than the destination.
The glissando motion process is shown in Figure~\ref{fig:glissando}.

After all glissandi are complete, we insert a brief gap in which endpoints appear as dim, steady markers, showing that the string texture has arrived and is holding its final pitch state.
The visualization is then processed into the five energy blocks: points are grouped into five strata along the vertical dimension and activated as dense, colored fields that organize moments of relative stasis.
Through the ``Block'' section, the endpoints remain highlighted and pulsed, presenting the idea that the blocks unfold while the string process remains anchored at its accumulated endpoint state, as shown in Figure~\ref{fig:energy}.

In parallel with the glissandi and energy block phases, we activate a reduced subset of sampling points as discrete brass and woodwind events.
These appear as small, color-coded point triggers whose onsets follow the beat-quantized onset equation (Eq.~\ref{eq:onset_quantization}), forming a sparse secondary layer that contrasts with the continuous string texture, as shown in Figure~\ref{fig:glissando} and Figure~\ref{fig:energy}.
\begin{equation}
\text{onset\_beats}
= \operatorname{round}\!\left(
      \frac{y_{\text{norm}} \cdot \text{total\_beats}}{\text{grid}}
   \right)\cdot \text{grid}.
\label{eq:onset_quantization}
\end{equation}

During the animation, we maintain a slow and continuous camera rotation at a fixed elevation, allowing the three-dimensional organization of the Pavilion to be perceived without interrupting temporal continuity.
The whole real-time visualization is shown in Fig~\ref{fig:geometry_animation} and\ video output.\footnote{Real-time visualization video output is available at: \url{https://youtu.be/zSj_I4n7Yqg}}

\section{Conclusion}\label{sec:conclusion}
This study proposes an architectural sonification system focused on the \textit{Philips Pavilion} created by Xenakis's \textit{Metastaseis}.
Through parametric geometric analysis, we revisited the generative logic behind Xenakis's transition from glissando structures to ruled surfaces in \textit{Metastaseis}.
We then reversed this process to sonify the \textit{Philips Pavilion}'s geometry.
Using parametric modeling, we reconstructed the Pavilion as nine ruled surfaces and extracted their governing ruling lines.
We subdivided each surface into interpolated structural lines and selected four evenly spaced lines per surface as the backbone of the string section, yielding 36 line segments that drive continuous string glissandi.
In addition, we sampled spatial points across the ruled surfaces and organized these point distributions into two complementary layers, comprising five energy blocks stratified along the vertical dimension to organize moments of stasis after the glissandi, and a reduced subset of points that triggers sparse brass and woodwind events to strengthen structural moments within the timeline.
By adopting the instrumentation associated with \textit{Metastaseis}, our system renders the Pavilion as a time-based musical form, allowing architectural structure to be perceived through motion, accumulation, and event-driven contrast.
To more clearly visualize our architecture sonification system, we also developed a real-time visualization platform. The Pavilion appears as a low-opacity point cloud with inactive structural lines, while musical events activate geometric elements in sequence.
Colored traversals reveal the unfolding of string glissandi along architectural trajectories; dense point activations visualize the energy blocks; and discrete point triggers mark brass and woodwind events.
Together, the audiovisual system frames architectural geometry not as a static outcome but as an active generative instrument whose structure can be heard and viewed through time.

Our study not only reinterprets Xenakis’s music-architecture process in reverse, but also suggests a new direction for exploring the auditory dimension of architectural design and interactive performance.
With contemporary parametric tools, architecture can be viewed as a dataset for sonification. Similarly, sonification gives architects a new design perspective beyond visual intuition.
Thus, architects can uncover structural logic, rhythmic relationships, and form potential at the auditory level, which were obscured in traditional geometric presentations.

However, there are several limitations in our study. From an architectural data perspective, the ruled surfaces were extracted using uniform interpolation and equivalent division of sampling points to ensure continuity in sonification. However, the ruled surfaces are complex and fixed-density structural lines and sampling points struggle to capture local curvature variations, which can weaken and simplify certain geometric features in the sonification results.
From a sonification perspective, although we have restored the instruments used in \textit{Metastaseis}, applying the same instrumental system universally cannot fully express the distinctive features of other architectural styles or explore broad musical possibilities.
From an interaction perspective, our sonification system visualized audiovisual elements but lacked user experience features common to interactive architectural systems.

In future work, we aim to explore further possibilities for architectural mapping strategies, extend the sonification system to more architectural cases, and enable the audiovisual representation of the \textit{Philips Pavilion} to interact with users in immersive environments such as VR/AR. Therefore, users can hear the architectural sonification results as they navigate the space, deepening the experiential connection between architecture and music.

%\begin{acknowledgments}
%At the end of the Conclusions, acknowledgments to people, projects, funding agencies, etc. can be included after the second-level heading ``Acknowledgments'' (with no numbering).
%\end{acknowledgments} 

%%%%%%%%%%%%%%%%%%%%%%%%%%%%%%%%%%%%%%%%%%%%%%%%%%%%%%%%%%%%%%%%%%%%%%%%%%%%%
%bibliography here
\bibliography{icmc2026_paper_template}

\end{document}